\documentclass[aip,apl,amsmath,amssymb,
preprint,%
]{revtex4-1}

\usepackage{color}

\usepackage{graphicx}
\usepackage{dcolumn}
\usepackage{bm}

\begin{document}

\title{Optical-resolution photoacoustic microscopy by use of a multimode fiber}

\author{Ioannis N. Papadopoulos}
\thanks{These authors contributed equally.}
\affiliation{School of Engineering, \'Ecole Polytechnique F\'ed\'eral de Lausanne (EPFL), Station 17, 1015 Lausanne, Switzerland}
\author{Olivier Simandoux}
\thanks{These authors contributed equally.}
\affiliation{Institut Langevin, ESCPI ParisTech, CNRS UMR 7587, INSERM ERL U979, 1 rue Jussieu, 75005, Paris, France}
\author{Salma Farahi}
\affiliation{School of Engineering, \'Ecole Polytechnique F\'ed\'eral de Lausanne (EPFL), Station 17, 1015 Lausanne, Switzerland}
\author{Jean Pierre Huignard}
\affiliation{Jphopto-consultant, 20 rue Campo Formio, 75013, Paris, France}
\author{Emmanuel Bossy}
\affiliation{Institut Langevin, ESCPI ParisTech, CNRS UMR 7587, INSERM ERL U979, 1 rue Jussieu, 75005, Paris, France}
\author{Demetri Psaltis}
\affiliation{School of Engineering, \'Ecole Polytechnique F\'ed\'eral de Lausanne (EPFL), Station 17, 1015 Lausanne, Switzerland}
\author{Christophe Moser}
\affiliation{School of Engineering, \'Ecole Polytechnique F\'ed\'eral de Lausanne (EPFL), Station 17, 1015 Lausanne, Switzerland}

\date{\today}

\begin{abstract}
We demonstrate Optical-Resolution Photoacoustic Microscopy (OR-PAM), where the optical field is focused and scanned using Digital Phase Conjugation (DPC) through a multimode fiber. The focus is scanned across the field of view using digital means, and the acoustic signal induced is collected by a transducer. Optical-resolution photoacoustic images of a knot made by two absorptive wires are obtained and we report on resolution smaller than 1.5$\mu$m across a 201$\mu$m by 201$\mu$m field of view. The use of a multimode optical fiber for the optical excitation part can pave the way for miniature endoscopes that can provide optical-resolution photoacoustic images at large optical depth.
\end{abstract}



\maketitle
Photoacoustic microscopy (PAM) is a rapidly evolving imaging technique that is capable of delivering multi-scale images based on the optical absorption properties of the investigated sample \cite{Li:2009bi}. Absorption of laser pulses by the sample, induces a local increase of the temperature generating an acoustic wave via the photoacoustic effect. An acoustic transducer is generally used to detect the generated acoustic wave enabling the formation of an image. One of the advantages of photoacoustic microscopy is that it can provide either label-free images of tissues, in which case the contrast rises from the intrinsic variation of optical absorption coefficient \cite{Zhang:2006cu, Maslov:2005ic}, or images of tissue based on exogenous contrast agents that can be used to target and image specific structures and metabolisms \cite{Agarwal:2007gk, Li:2008db, DeLaZerda:2008dx}.  Photoacoustic microscopy approaches can be divided in two categories, based on the image resolution. In acoustic-resolution photoacoustic microscopy (AR-PAM), the image resolution is dictated by the frequency response of the ultrasound detection device. With this approach, images with resolution ranging from millimeter to less than a hundred microns can be obtained at depths much larger than the ballistic range of optical propagation in scattering media, defined as the optical transport mean free path. At this depth, typically 1 mm in biological tissue, purely optical techniques are limited by multiple scattering \cite{Ntziachristos:2010gi}. The image reconstruction is based on acquiring ultrasound signals for different positions of the transducer, either by scanning a single element transducer or by using multi-element probes. In optical-resolution photoacoustic microscopy, the excitation light is focused into a diffraction-limited spot, usually with an objective lens. In this configuration the optical energy deposited on the sample, and therefore the region that generates the acoustic waves, are limited only to the illuminated diffraction-limited voxel, yielding photoacoustic images with optical resolution \cite{Maslov:2008fv, Zhang:2010ek, Xie:2009gg, Hajireza:2011dd}. The generation of full-sized images is based on the raster scanning of the optical excitation spot. However, as OR-PAM relies on the ability to focus light into the sample of interest, it is feasible only within the ballistic regime of optical propagation in scattering media, therefore generally limited to penetration depths smaller than 1mm for biological tissue.  

While the imaging depth of AR-PAM is limited by the so called hard limit of optical penetration, dominated by the absorption properties of the investigated tissue, the imaging depth of OR-PAM is limited by the optical transport mean free path. In order to overcome these limitations, endoscopic modalities have been deployed for the acquisition of photoacoustic images deep inside tissue. Similar to conventional optical endoscopy, fiber bundle endoscopes have been adapted to add photoacoustic imaging capabilities \cite{Shao:2012cn}. A different approach of a purely photoacoustic endoscope has also been proposed \cite{Yang:2009di, Yuan:2010jn} where a single mode fiber is used for the delivery of a diffused excitation optical field and the integrated transducer picks up the acoustic signal. These devices are equipped with a rotational motor so that the excitation and detected field can be scanned around the endoscope's axis. The proposed devices are still limited in terms of resolution, 58$\mu$m lateral resolution and the smallest achieved diameter of the endoscopic head was 2.5mm \cite{Yang:2012ct}.

Fiber bundles for purely optical endoscopes, where each of the single mode fibers of the device acts as a single pixel of the final image, dominate the commercial domain, however the resolution is limited by the distance between adjacent fiber cores (usually $\sim 5\mu$m) \cite{Gmitro:1993gi}. Moreover, designs that combine optical fibers, conventional focusing optical elements and mechanical actuators have been used in versatile high resolution endoscopes \cite{Bird:2003ky}. Yet these designs remain relatively large (larger than 2mm). Recently, different groups have explored the possibility of building functional ultra-thin imaging devices based solely on multimode fibers exploiting the large number of degrees of freedom available in these waveguides. Fluorescent and linear scattering imaging has been demonstrated \cite{Bianchi:2012du, Cizmar:2012gz, Choi:2012dx, Papadopoulos:2013uka, Mahalati:2013uu}.

In the present paper we demonstrate an optical-resolution photoacoustic imaging modality in which a multimode fiber is used as the source of the optical excitation field. We use Digital Phase Conjugation (DPC) to focus and digitally scan a diffraction-limited spot at the distal tip of the multimode fiber \cite{Papadopoulos:2012ko} therefore eliminating the need for optical lenses and mechanical actuators. The small diameter of multimode fibers along with their ability to focus light into a micron-sized spot, pave the way for the implementation of optical-resolution PAM deeper than the ballistic range of light propagation. In the proposed configuration, a needle-type endoscope \cite{Papadopoulos:2013uka} can be used to  bring the optical excitation directly against the sample of interest by penetrating the tissue. The generated acoustic signal can then be picked up by a transducer placed on the surface of the tissue. The demonstrated implementation could enable the fabrication of ultra-thin, minimally invasive endoscopes capable of multi-modal imaging (linear scattering, fluorescence, optical absorption) deep inside tissue. 

However, the combination of large number of degrees of freedom and small size in multimode fibers does not come without a cost. The optical field that propagates through the fiber couples to the different modes and reaches the output where it appears scrambled. In order to undo the modal scrambling induced by multimode fiber propagation, we use Digital Phase Conjugation (DPC)  \cite{Bellanger:2008gs,Cui:2010wr,Hsieh_digital_phase,Papadopoulos:2012ko}. Optical phase conjugation has been used to undo the scattering induced from turbid media where the conjugate field back-propagates through the medium retracing its way through, finally generating the original image \cite{Leith:1966ki,Kogelnik:1968uy,Yaqoob:2008dt}. In the digital version of phase conjugation, the photorefractive crystal is substituted by a digital sensor and a Spatial Light Modulator (SLM), therefore highly increasing the versatility of the system.

\begin{figure}
\includegraphics{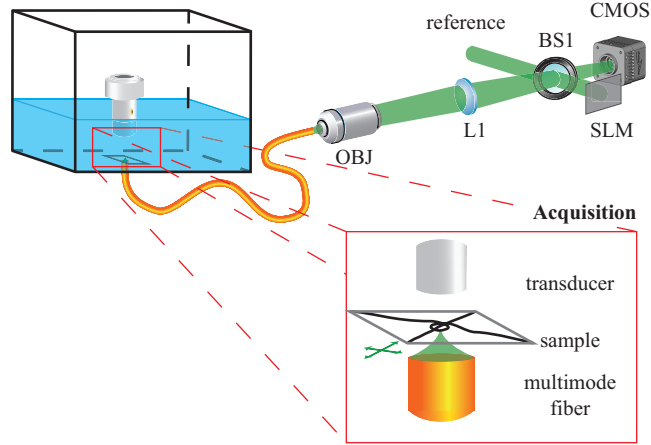}
\caption{\label{fig:1} Schematic of the experimental setup for the generation of optical-resolution photoacoustic microscopy images using a multimode fiber for the optical excitation. After the initial calibration of the system so that it can perform the focusing and scanning of the laser light as described in reference \cite{Papadopoulos:2012ko}, the distal tip of the fiber and the transducer were immersed in water. The reference beam is modulated by the SLM forming the optical phase conjugate beam. The field with the engineered wavefront, enters the fiber and after propagating through, forms a sharp focus spot at the desired position. When the light is focused on an absorbing feature, photoacoustic waves are emitted and detected by the acoustic transducer. By digitally raster scanning the focused spot and acquiring the data from the transducer, an optical resolution photoacoustic  image is generated.}
\end{figure}

Figure 1 presents a schematic of the experimental setup for OR-PAM implementation using a multimode fiber for the excitation. The initial step in the process is the calibration of the multimode fiber so that a focused spot can be digitally raster scanned across a designed field of view. The output of a Q-switched Nd:YVO4 laser (wavelength 532nm and pulse width of 5ns, NL-201, EKSPLA) is focused by a 40x, 0.65 NA microscope objective (not shown in Fig. 1) on a diffraction-limited spot, 250$\mu$m away from the distal tip of the fiber and then propagates through the multimode fiber (220$\mu$m diameter, 0.53NA multimode fiber, Ceramoptec) reaching the proximal tip where it generates a speckle pattern. The speckle pattern is imaged on a CMOS detector through a 4-f imaging system and is combined with the reference beam to generate an interference pattern. The phase of the output field is calculated from the digital reconstruction of the captured hologram. The calculated phase is assigned onto the phase-only SLM modulating the reference beam and finally generating the optical phase conjugate beam. The phase conjugate beam back-propagates through the system and because of the tailored phase, the scrambling is undone, generating a sharp focus spot at the initial position. Scanning the calibration focus spot and calculating the respective phase as described in reference \cite{Papadopoulos:2012ko}, we can generate a phase look up table that will allow us to scan the focused spot across a regular orthogonal grid. The whole process of the fiber calibration is done without the presence of water to avoid any disturbance. We generate spots of 1.2 $\mu$m, 250$\mu$m away from the distal tip and we scan 134 by 134 spots that are 1.5$\mu$m away from each other, therefore yielding a square field of view of 201$\mu$m by 201$\mu$m. The pulse energy at each focus spot was estimated at 20nJ/pulse.

During the imaging part of the experiment, we place a transducer (20 MHz, spherically focused, 12.7mm focal distance, 6.3mm diameter, P120-2-R2.00, Olympus, Japan) in front of the fiber as shown in the inset of Fig. 1 and fill the area with water. The transducer remains still during the whole photoacoustic experiment. The presence of water will affect the working distance but not the distance between adjacent focused spots. The sample used for the experiment consists of two black nylon wires (20-30$\mu$m nominal diameter) tied into a knot and placed on a sample holder. The knot is placed at the working distance of the fiber, which is defined as the focusing distance away from the distal fiber tip. The focused spots are scanned across the regular grid one by one. For each position of the focused spot, the transducer picks up the generated acoustic wave and converts it to an electric signal which is amplified by a low noise amplifier (DPR500, remote pulser RP-L2, JSR Ultrasonics, USA) and displayed on an oscilloscope. A computer is used to synchronize all the required actions by sequentially projecting the phase on the SLM for the generation of the focus, then acquiring the data from the transducer through the oscilloscope and repeating these steps for the scanning and acquisition of the image.
\begin{figure}[htb]
\includegraphics[scale=1]{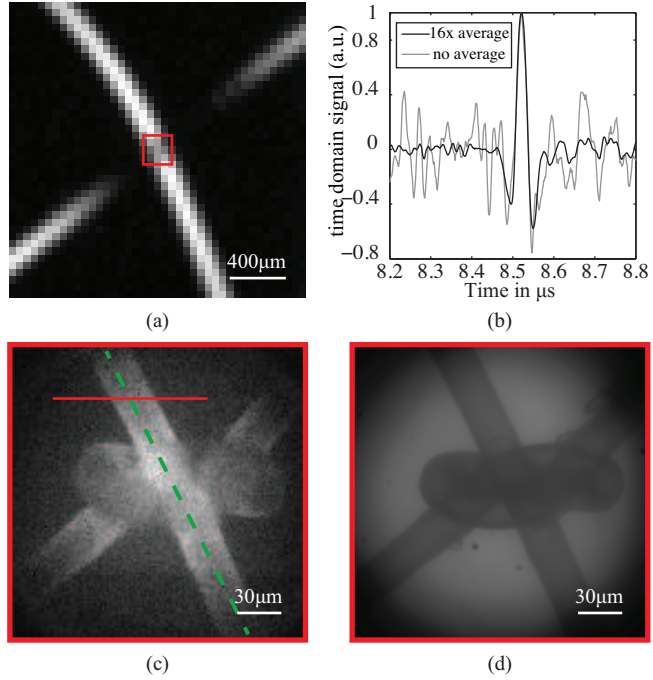}
\caption{\label{fig:2} (a) Pulse-echo ultrasound image by raster scanning the transducer. The resolution is determined by the central frequency and the aperture of the transducer. The ultrasound image resolution cannot render any detail of the tied wires in the center. (b) Time-domain acoustic signal captured by the transducer when a focus spot is projected at the center of the knot. The acoustic signal without averaging is shown in grey and the signal obtained by averaging 16 times is presented in black. The SNR of the non-averaged signal is 6 and is raised to 25 by 16 times averaging. (c) Optical-resolution photoacoustic image. Digital phase conjugation is used to render and scan a tight optical focus, 250$\mu$m away from the fiber. The image is generated by calculating the energy of the acoustic signal captured from the transducer at each spot. No mechanical scanning is employed at any step of the photoacoustic imaging process. Each pixel corresponds to 1.5$\mu$m and the total field of view is 201$\mu$m by 201$\mu$m. (d) White light optical image of the knot immersed in water using a 20x, 0.35 NA microscope objective showing the same field of view as in (c). The OR-PAM image captured with the use of the multimode fiber reveals the details of the knot. Images in (c) and (d) correspond to the red square in (a). }
\end{figure}

In Fig. 2, we demonstrate and compare the results of the different imaging modalities implemented. In Fig. 2a, we present the ultrasound pulse-echo image, captured by sequential raster scanning of the transducer. In this modality, the resolution is defined as for acoustic-resolution photoacoustic microscopy, by the central frequency and the aperture of the transducer. The calculated 2w$_0$ parameter of the transducer is ~200$\mu$m and the effect of this can be clearly seen on Fig. 2a where the 30$\mu$m wide wires appear as thick structures and no detail of the knot can be resolved. The pixel size of the image in Fig. 2a is 50$\mu$m. Figure 2b presents the time-domain acoustic signal captured by the transducer when the light emerging from the multimode fiber is focused on the wires. The grey line corresponds to the raw electric signal from the transducer, while the black line is the result of 16 times averaging. The non-averaged signal has a SNR of 6, which rises to 25 when 16 times averaging is performed. The SNR level of the non-averaged signal is sufficient to allow the acquisition of images with just a single pulse. By scanning the focused spot and measuring the energy of the acoustic signal at each scanning position, the OR-PAM image shown in Fig. 2c is obtained. The image in this case is acquired by averaging 16 times and each pixel corresponds to 1.5$\mu$m. Comparing the OR-PAM image with the white light optical image shown in Fig. 2d, we see that the OR-PAM image, generated with the use of the multimode fiber for the excitation, gives a high level of detail closely resembling the white light optical image. The imaging plane is set exactly at the plane of the wires and the rendered 2-d image illustrates the 3-d structure of the knot.
\begin{figure}
\includegraphics{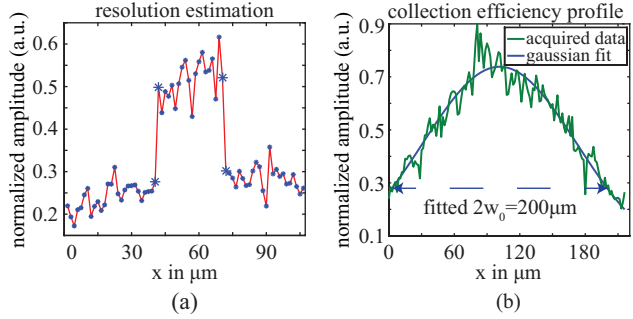}
\caption{\label{fig:3} The cross-sectional plots shown in (a) and (b) correspond to the solid red and dashed green lines respectively of Fig. 2c. (a) As demonstrated by the transition between the background to the signal coming from the wire, the resolution of the system has an upper limit of 1 pixel, therefore 1.5 $\mu$m. (b) The cross-sectional plot along the wire shows that the collected signal is weaker towards the edges of the image (green dashed line). A Gaussian fit of the acquired data has a 2w$_0$ parameter of 200$\mu$m, equal to the 2w$_0$ parameter of the focused transducer. The difference in the signal level is therefore attributed to the Gaussian-like profile of the collection efficiency of the transducer used.}
\end{figure}

The resolution of the system can be estimated by taking a cross-sectional plot along the wires and measuring the distance between the background and the wire signal. To avoid the need to interpolate the acquired data, which would cause the generation of artifacts, the cross-sectional plot is drawn horizontally across the raster scanning, not exactly perpendicular to the wires therefore resulting in an underestimation of the achieved resolution. Figure 3a shows the cross-sectional plot along the solid red line in Fig. 2c while Fig. 3b is the cross-sectional plot along the dashed green line. As can be seen in Fig. 3a the transition between the signal of the background to the signal from the wire, occurs within 1 pixel of the image (1.5$\mu$m) defining therefore the upper limit of the resolution to 1.5$\mu$m. To characterize the collection efficiency of the system, we draw the cross-sectional plot along the wire as shown in Fig. 3b. We can see that the collected signal is weaker towards the edges of the image. Performing a curve fitting with a Gaussian profile, we see that the fitted curve has a 2w$_0$ parameter equal to 200$\mu$m, which corresponds well to the calculated 2w$_0$ parameter of the focused transducer. Therefore the different levels of the collected signal between the center and the edges of the image are attributed to the Gaussian collection efficiency of the focused transducer. 

The phase look-up table calculated for the focusing and scanning of the light field after the multimode fiber is valid only for a specific conformation of the fiber. In case the fiber changes conformation (bending) the phase look-up table has to be recalculated. In order to overcome this restriction, the proposed device should be implemented by keeping the multimode fiber in a rigid jacket. The small diameter of the fiber along with the absence of any lenses or mechanical actuators means that the whole rigid head can be made very small and in principle smaller than 500$\mu$m. This enables the generation of needle-type devices, in which the optical excitation is brought against the interrogated sample by directly penetrating through the tissue. The transducer can be placed against the tissue surface to pick up the generated acoustic waves. The main limitation in this geometry arises from the attenuation of ultrasound through the tissue during the signal collection stage. The attenuation of ultrasound in tissue increases linearly with frequency, on the order of 0.5 dB/cm/MHz, limiting the detection depth to a few mm for ultrasound frequencies of several tens of MHz. In the context of AR-PAM, imaging depths as large as 3mm have been demonstrated for a 50MHz transducer by Zhang et al.~\cite{Zhang:2006cu}. For the approach proposed here, optical detection of ultrasound directly at the tip of the fiber would overcome the limitation caused by the remote detection of ultrasound. In conclusion the scheme demonstrated in this work could trigger the development of endoscopes capable of delivering optical-resolution photoacoustic imaging far beyond the ballistic regime of light propagation. Within a single approach, the multimode fiber system could deliver multi-modal images based on fluorescence, linear scattering and optical absorption properties of the sample of interest.

\end{document}